# Seeding Life on the Moons of the Outer Planets via Lithopanspermia

R. J. Worth[123], Steinn Sigurdsson[123], Christopher H. House[24]

**Abstract**

Material from the surface of a planet can be ejected into space by a large impact, and could carry primitive life forms with it. We performed n-body simulations of such ejecta to determine where in the Solar System rock from Earth and Mars may end up. We find that, in addition to frequent transfer of material among the terrestrial planets, transfer of material from Earth and Mars to the moons of Jupiter and Saturn is also possible, but rare. We expect that such transfer is most likely during the Late Heavy Bombardment or during the next one or two billion years. At this time, the icy moons were warmer and likely had little or no icy shell to prevent meteorites from reaching their liquid interiors. We also note significant rates of re-impact in the first million years after ejection. This could re-seed life on a planet after partial or complete sterilization by a large impact, which would aid the survival of early life during the Late Heavy Bombardment.

---

1   The Pennsylvania State University Dept. of Astronomy & Astrophysics, University Park, Pennsylvania
2   Penn State Astrobiology Research Center, University Park, Pennsylvania
3   Center for Exoplanets and Habitable Worlds, University Park, Pennsylvania
4   The Pennsylvania State University Dept. of Geosciences, University Park, Pennsylvania

# 1. Introduction and Background

Panspermia is the hypothesis that life can be spread between planets and planetary systems. One class of panspermia is lithopanspermia, in which pieces of rock are the mechanism for dispersal (Tobias & Todd, 1974; Melosh, 1988). Rock fragments can be ejected from an inhabited planet's surface via large meteor impact. This ejected material can then travel through space and may land on another planet or moon, as we have seen in identified meteorites from Mars found on Earth (Bogard & Johnson, 1983; Carr *et al.*, 1985). If an ejected rock encases sufficiently resilient organisms, life could be seeded on its destination planet or moon.

*1.1 Characterizing impacts and ejections*

Of the over 53,000 meteorites found on Earth, 105 have been identified as Martian in origin (Meteoritical Society, 2012[*]). The oldest of these observed falling to Earth is dated from 1815, though many have been discovered after falling. Some of our samples are compositionally similar, and may have come from a common impact. The entire collection of Martian meteorites represents at least four impacts on Mars (Nyquist *et al.*, 2001).

One of these meteorites, Allen Hills 84001 (ALH84001), was initially claimed to contain evidence of life in the form of poly-aromatic hydrocarbons (PAHs), carbon globules, magnetite crystals, and fossilized nanobacteria (McKay *et* al., 1996). This claim sparked recent investigation into quantifying and characterizing lithopanspermia. Each of the claimed biomarkers has been shown to be reproducible inorganically (see Martel *et al.* 2012 for a review, and references therein for specific counterarguments, e.g. Cisar *et al.*, 2000; Benzerara *et al.*, 2003; and Golden *et al.*, 2004). Despite the lack of hard evidence for life existing in any of the transferred rocks, the debate motivated a number of important studies of the process of lithopanspermia, which will be discussed below.

It is known that rock can be exchanged between planets; therefore, if life can survive the transfer, it is probable that life from Earth has already been brought to other planets. Also, it is theorized that Mars may have been habitable during the Noachian period, when the young planet is thought to have been warm and wet (McKay *et al.*, 1992). Carbonates in the Allen Hills meteorite mentioned above precipitated in a fluid of 18°C about 4 billion years (4 Gyr) ago, when the planets were only 500 million years old (Halevy *et al*., 2011). Early Mars appears to have been warm and wet, providing prime conditions for the development of life. Thus, life from Earth or Mars could have been brought to other planets in the solar system, and any habitable niches may continue to harbor descendants of such life. It is even possible that life on Earth originated on Mars (McKay *et al*., 1996).

---

[*] http://www.lpi.usra.edu/meteor/metbull.php

*1.2 Previous Dynamical Studies*
For decades, astronomers have been attempting to model the dynamics of ejected rocks migrating about the Solar System, but technological limitations have always restricted the scope of such simulations. Early studies used Arnold-Öpik integrators (Melosh & Tonks, 1993). These methods underestimated the amount of interplanetary exchange and overestimated the timescale of transfer because they did not include secular effects, in which the resonance of precession frequencies causes increases in eccentricity and inclination in small bodies on timescales of Myr. Without these effects, transfer between Earth and Mars was still observed, but it took place very slowly, transferring only a few percent of ejecta over 150 Myr.

    Subsequent studies were conducted by Gladman & Burns (1996), Gladman *et al.* (1996), and Gladman (1997). Their use of symplectic integrators improves accuracy and greatly increases the speed at which interplanetary transfer occurs in the simulations. The difference between methods is discussed in more detail in Dones *et al.* (1999), but in brief, the inclusion of secular effects results in gradual eccentricity increases, which causes objects initially originating from and orbiting near one planet to eventually cross the orbit of another within timescales of a few million years. At this point, the meteoroid may impact the other planet or be scattered into many possible new orbits.

    Comparing transfer times at different ejection velocities with the ages of collected meteorites, Gladman *et al.* (1996) determine that most rocks are ejected at velocities just above the escape velocity. Additionally, they found that most transfer between planets occurs on the timescale of <10 Myr, after which most remaining meteoroids ultimately fall into the Sun. The Gladman studies included all the planets from Venus through Neptune, and particles were considered ejected from the Solar System if they went beyond 100 AU from the Sun. No collisions with the outer planets or moons were observed, due to the relatively small number of objects in the simulations.

    Gladman (1997) found that some objects were held within the asteroid belt between Mars and Jupiter by secular resonances for millions of years. This timescale is sufficient for these objects to be collisionally destroyed, which reduces interplanetary transfer by a small amount. However, most rocks subjected to this fate were long-lived (>10 Myr) objects that would have ultimately ended up in the Sun, and thus are not significant to our study.

    A comprehensive, multidisciplinary study of transfer between Earth and Mars (Mileikowsky *et al.*, 2000) combined studies of ejection rates, dynamical transfer, biological survival requirements, and impact physics to determine the amount of viable biological material transferred between Earth and Mars. Survivability constraints are based on the microbe species *Bacillus subtilis* and *Deinococcus radiodurans* R1. The authors calculate the number of rock fragments ejected that satisfy two survivability criteria: remaining below

100°C through the ejection, and being large enough to shield microbes from the harms of a space environment for the duration of travel. They estimate that a rock of 3 m across shields *D. radiodurans* for 10 Myr, and 3.3 Myr for *B. subtilis*.

Mileikowsky *et al.* (2000) estimate the total number of ejected fragments from Earth and Mars over the roughly 4 Gyr since life arose on Earth. In order to update these values we repeat their calculations using updated cratering rates (Ivanov & Hartmann, 2007) to estimate the numbers of viable fragments ejected. We find that, over the last 3.5 Gyr, approximately $3 \times 10^8$ suitable fragments have been ejected from Earth, and $6 \times 10^8$ from Mars.

Gladman *et al.*, (2005) perform simulations in which 36,000 particles were ejected from Earth with $v_\infty$ (residual velocity, or velocity when the object is an infinite distance from other bodies) from 0.67 to 12.41 km/s and integrated over 30 kyr. Residual velocity is defined as
$$v_\infty^2 = v_{ej}^2 - v_{esc}^2$$
where $v_{ej}$ is the ejection velocity relative to the planet, and $v_{esc}$ is the escape velocity
$$v_{esc}^2 = 2GM_{planet}/r$$
The authors found transfer to Venus (0.1%), back to Earth (1%), and to Mars (0.001%) in the 30,000 years of this study, which represents only the very beginning of our time frame.

Gladman *et al.* (2006) make use of technological improvements to expand the scope of their simulations. Using 18,000 particles in orbits evolved over 5 Myr, with $v_\infty$ of 5 to 10 km/s, they used the number of fly-bys to extrapolate the number of collisions that might occur on Titan and Europa, and conclude that the rate of transfer would be on order of ten objects per 600 million ejected, or roughly $10^{-6}$ %.

A more recent study (Reyes-Ruiz *et al.*, 2012) built on the foundations of Gladman *et al.* (2005), again studying high-velocity ejections from Earth on shorter timescales (30 kyr). They found some transfer to Jupiter, but only at very high ejection velocities ($v_\infty$ = 9.64 and 12.1 km/s). The velocities used here and by Gladman *et al.* (2006) are significantly higher than what Gladman *et al.* (1996) predicts to be typical, but result in faster transfer and thus less expensive simulations.

## 2. Methods

Using the hybrid symplectic integrator in MERCURY (Chambers, 1999), we simulated the ejection of tens of thousands of meteoroids from both Earth and Mars. Because they are so much smaller than the other bodies, the meteoroids were treated as test particles with no gravitational influence on one another or on the planets in the system. The program integrated over 10 Myr, with one-day timesteps. All eight planets were included in the simulations.

The inputs for the simulations were the positions and motions of the planets of the Solar System – taken from the Jet Propulsion Laboratory online ephemerides system HORIZONS (JPL) – and positions and motions of the meteoroids. The meteoroids were placed at a distance near one Hill radius away from their planet of origin at randomly-generated angles, thus covering the surface of the Hill sphere centered on the planet. This radius is roughly

$$R_H \approx a\, (M_{planet}/3M_{Sun})^{1/3}$$

For Earth, objects were placed at $2 \times 10^{11}$ cm or 360 Earth radii, while for Mars the distance was $1 \times 10^{11}$ cm or 300 Mars radii.

The ejected rocks were given the origin planet's orbital velocity plus a strictly radial ejection velocity vector varying randomly between one and three times the escape velocity from the planet at that location, simulating ejection. At these distances, escape velocities are 0.58 km/s for Earth and 0.28 km/s for Mars. The residual velocity $v_\infty$ ranges from almost zero to a maximum of 1.6 km/s for Earth and 0.78 km/s for Mars. The flat distribution of velocities over this range does not represent a natural distribution that would result from a given impact, but allows us to explore potential dependencies of impacts on the initial velocity across a range of energies.

Moons were not included in the main simulations, as they are near enough to their host planets to be considered close encounters by the integrator, thus continuously triggering the switch to the non-symplectic integrator, which is an order of magnitude slower. To estimate impact on moons for ejecta on orbits with essentially random orbital phases, rates can be estimated from the geometric projected area of the target moon, corrected for gravitational focusing, an effective application of Liouville's theorem, as noted by Melosh (2003). It is possible that impact rates can be enhanced by ejecta being captured into orbits bound to the planet and going through multiple passes through the system of moons on chaotically resonant orbits, a phenomenon well known in 3-body dynamics (Hut and Bahcall, 1983). The only way to test whether this is a signficant effect is through numerical simulations.

To calculate the probability that an object may hit a moon instead of a planet, we performed many short simulations of meteoroids approaching Jupiter and Saturn with the moons included, and counted the number of times they hit either the planet or one of the moons. These simulations were again performed with MERCURY[*], including all the Solar System bodies as well as the moons of interest as "big" objects, and the meteoroids as "small" ones. Since the moons are within the close-encounter cutoff radius, the simulation proceeds using a slower but more precise variable-timestep integrator. For the meteoroids' trajectories, we put them near the planet's orbit at a random orbital phase, adding random variations of position up to $10^{11}$ cm, and of velocity up to 100 cm/s. This procedure was repeated hundreds of times, with different random orbital positions of the moons, until we could estimate a ratio of moon-to-planet impacts to apply to the planet-to-planet transfer rates.

---

[*] We noted a problem with the selection criteria for collisions with the central body when satellite bodies were involved. Changes to the code were made to allow stable moon orbits. For a copy of the code, contact R. J. Worth at rjw274@psu.edu.

Calculating a 10 Myr trajectory took approximately 1,000 CPU seconds per ejected object, or roughly 1,000 CPU days for the entire population of about 100,000 objects. Most of the simulations were performed with 2.6 GHz Intel Xeon cores. This only counts the time used on the main simulations, and not the smaller moon-impact simulations. The absolute amount of CPU time used was substantial, as each trajectory must be modeled for many dynamical times, and the code requires a dedicated processor over long periods of time. However, this amount of time is modest by modern standards, and breaking the population of small objects up into several simulations allows us to run in parallel on multiple processors, greatly reducing the wall time.

## 3. Results
### 3.1 Total rates of transfer between planets
The majority of collisions by far were onto the planet of origin for the simulation (see Table 1). Large numbers also impacted the planets inward from the origin planet, and many meteoroids were either captured by the Sun or ejected from the system entirely. Small numbers were gradually perturbed into orbits allowing them to impact planets outward from their origin, namely from Earth to Mars, Jupiter, and Saturn, and from Mars to Jupiter. In general, the number of impacts decreased with distance from planet of origin, except for the Sun and ejections from the system, which serve as outer boundaries and catch large numbers of objects. In general, the planet of origin influences the rate of transfer most strongly for collisions on Venus, Earth, and Mars, while for more distant destinations, the importance of which planet the object was ejected from is reduced.

### 3.2 Timescales of transfers
The distribution of impact times can be seen in Fig. 1. The majority of collisions occurred as ejected meteoroids re-impacted their planet of origin on short (<0.5 Myr) timescales. These impacts drop off exponentially with time, while impacts on more distant bodies occur later and less frequently, with peaks at later times for destinations more distant from the planet of origin, as shown below in Table 2. The distributions for some bodies appear to be truncated by our simulations' lengths of 10 Myr, indicating that extending the length of simulation would increase the proportions of impacts on these planets.

### 3.3 Impacts at 10-30 Myr

Our primary interest for these simulations is in whether ejected material could carry life between Solar System bodies. Based on the estimates of Mileikowsky *et al.* (2000) that moderately-sized ejecta could shield organisms on timescales of millions of years, and the findings of Gladman *et al.* (1996) that most transfer occurs in the first 10 Myr of their simulations, we set 10 Myr as the length of most of our simulations. However, for a small number of objects (9,000 from Earth, 6,000 from Mars) we extended the simulations to 30 Myr, to see how transfer rates might change on longer timescales.

The small number of objects in these simulations means the results are noisy, especially for transfer to the outer planets where the numbers are low to begin with. We find that impact rates after 10 Myr decline for most destinations, with a few exceptions: rates of collision with the Sun or ejection from the Solar System peak between 10 and 15 Myr from initial ejection; transfer rates from Earth to Jupiter continue until about 15 Myr from ejection, and from Mars to Jupiter until nearly 25 Myr; and transfer from Mars to Mercury and Venus increases after the first 10 Myr. However, the numbers for Jupiter, Mercury, and Venus are particularly small and subject to large amounts of noise, so the significance of these results is uncertain. The median transfer times for these simulations can be found in Table 2.

*3.4 Moon impact rates*

We performed additional simulations of meteoroids approaching Jupiter and Saturn with the moons included, as described in section 2. These simulations included either two of Saturn's moons (Titan and Enceladus), or Jupiter's four Galilean moons (Io, Europa, Ganymede, and Callisto), as well as a set of randomized meteoroids already near the giant planet. We ran these small simulations many thousands of times to calculate the ratio between the number of meteoroids that hit the planet to the number that hit each moon. We then used this to estimate how many of the planet impactors from the main simulations would actually hit the moons instead, had they been present.

We found that many of the moons were hit more often than the Liouville theorem (discussed in Section 2) would predict, as shown in Fig. 2a. The exceptions to this were Titan and Callisto, which both received fewer impacts than expected. The impacts recorded in these simulations can be found in Table 3.

We observed a correlation between the moons' impact ratios and a combination of their parameters, as illustrated in Fig. 2b. The moons and planets in our study indicate that the collision ratio between a satellite and parent body is proportional to the ratios of their gravitational focusing factors, times the planet's Hill sphere radius divided by the square of moon's semimajor axis. If the correlation proves applicable over a wide range of satellite systems, it could prove a useful tool.

# 4. Discussion
## *4.1 Correlation with initial conditions*

Using the statistical program R (R Core Team, 2013), we attempted to quantify which parameters were significant in determining the destination and timescale of the meteoroids' collisions. We used the function *glm*, which calculates a general linear model for a data set. The user chooses a response variable $Y$ (the parameter we are trying to fit) and one or more predictor variables $X_i$ (the parameters to which we want to fit the response variable). The program uses maximum likelihood estimates to fit the response as a function of the predictors, as described in more detail in the Appendix. It also determines the significance of each parameter based on a false alarm probability $P$, the probability that the given results could occur randomly. Convention dictates that $P < 0.05$ indicates significance.

Below, we discuss two fits: destination as a function of planet of origin, initial velocity $v_0$, and two ejection angle parameters ($\varphi$ indicates the direction within the ecliptic plane, while $\theta$ indicates the direction above or below the plane); and collision time as a function of destination, origin, initial velocity, and ejection angle.

We fit the destination by using it as the response variable in the model, with the initial conditions of ejection (planet of origin, initial velocity, and two angles parameterizing the direction of ejection from the planet) as predictors. The false alarm probability for the origin parameter is smaller than the program can calculate (i.e. less than the minimum nonzero double-precision floating point value, $2.225074 \times 10^{-308}$), indicating (unsurprisingly) that the planet of origin is extremely important in determining which planet an ejected object will collide with. The initial ejection velocity is also quite important in determining whether an object will collide with its planet of origin, but otherwise is not statistically significant. The direction of ejection is insignificant except for one case: a high polar angle has a small but statistically significant influence on the chances of re-impact. This seems to indicate that ejection angle and velocity are randomized before most objects impact a planet.

Using the same program, we then fit the impact time as a function of destination and the same four initial properties used above. We found that the destination is the most significant predictor of impact time, and origin is once again also very significant. The initial velocity is also statistically significant, but this is likely due to the correlation between velocity and likelihood of re-impacting, which occurs at far earlier times than other impacts. The ejection angle parameters are both insignificant.

Thus, overall we find that the fate of an ejected rock depends on the planet it is ejected from, and to some extent the velocity with which it is ejected, but otherwise is largely due to chance. Once an object gets beyond the influence of its origin body, its ejection parameters are no longer important.

*4.2 Calibrate with Mars-Earth transfer rates*

The rates of transfer from Mars to Earth and Earth to Mars found by Mileikowsky *et al.* (2000) are

$$f_{M2E} = 0.6 \% \times T$$
$$f_{E2M} = 0.016 \% \times T$$

for *T* up to 10 Myr, or 6% and 0.16% for the full length of our simulations. This is in reasonable agreement with our results of 0.2% for Earth-to-Mars transfer, but our 3% Mars-to-Earth transfer rate is only half of their value. Fig. 3 shows a stronger velocity dependence in this transfer, so these results appear consistent with our lower ejection velocities.

*4.3 Estimated numbers of transfers*

Mileikowsky *et al.* (2000) goes into detail estimating the number of rocks ejected from Earth and Mars with favorable conditions for propagating life. The fragments of ejecta must be of sufficient size to protect life for a journey through space of millions of years. They must also receive low enough shock pressures that the entire rock is not sterilized in ejection. We repeat the calculations described in sections 3.1 and 3.2 of Mileikowsky *et al.* (2000), but use the Martian cratering rates from Table 1 of Ivanov & Hartmann (2007). We scaled the ejected fragments by the transfer probabilities we found, giving us estimates of the total number of viable fragments that may have reached each destination. These results are presented in Tables 4 and 5.

The rates are provided for logarithmic bins of crater size, and we obtain a conservative estimate by using the minimum size in a given bin for subsequent calculations. We derived the impactor size *L* for each crater diameter *D*. For Earth, we scaled the Martian impact rate by the relative rates for Earth and Mars given in Table 4 of Ivanov & Hartmann (2007). Additionally, since the rates are given per $km^2$, we multiplied by the land surface area of the planet to obtain the whole-planet impact rates.

Mileikowsky *et al.* (2000) find that the approximate number of unsterilized fragments *n* (i.e. heated to no more than 100°C and subjected to pressures no more than 1 Gpa) is independent of *L*. For Mars impacts, $n = 2 \times 10^7$ and for Earth, $n = 6 \times 10^8$. We then calculated the fraction of unsterilized fragments that are at least 3 m in diameter, the minimum size able to shield organisms from the damages of space for up to 10 Myr. (For the benefit of readers who may wish to reproduce these calculations, we note that the paper mistakenly states that $F(2l) = 0.04$, when it is actually 0.4. Additionally, the equation relating transient to final diameter for Earth craters has 3 km as the first transition point, while the values in Table II imply that it should be 30 km.)

For each size *L,* we multiplied the impact rate over 3.5 Gyr by the number of unsterilized fragments of sufficient size. As stated in Mileikowsky *et al.* (2000), an impactor with a relatively high velocity ( >15 km/s on Mars or >30 km/s on Earth) is needed to eject a substantial number of fragments. They estimate only about 1/3 of incoming objects are fast enough, so we reduced the impact rates accordingly. Finally, we summed over all values of *L* to obtain the estimates for total viable fragments from each planet. Very large impactors with less than one expected high-velocity impact on land over 3.5 Gyr were excluded from the sum. We estimate roughly $2 \times 10^8$ qualifying meteoroids have been ejected from Earth and $8 \times 10^8$ from Mars over 3.5 billion years (this time span is both roughly the time since the Late Heavy Bombardment (LHB) and the period that Earth is known to have contained life).

As noted by Ivanov & Hartmann (2007), the eccentricity of Mars varies significantly, and the impact rates on its surface vary as a result. They show in Table 4 that the variance is within a factor of two over the known eccentricity range. We note that this effect could cause a moderate change in our estimation for Mars ejecta rates.

Generally, Earth's high surface gravity results in less mass being ejected, broken into smaller and more numerous fragments relative to an impactor of the same size on Mars. On Earth, a few very large impacts will eject huge numbers of viable fragments, while on Mars the same fragment size will be primarily produced by smaller impactors, in small but more frequent batches. Over 3.5 Ga, the largest portion of the ejecta from Earth come from impactors ~19 km in size, which will eject $1.6 \times 10^8$ viable fragments from a single impacts over the course of 3.5 Ga. On Mars, the most ejecta come from ~5 km impactors, which eject $2.7 \times 10^8$ such fragments, around 8 million each from approximately 35 impacts.

For the impacts under discussion, Mileikowsky *et al.* (2000) estimate that the mass ejected is $5.3 \times 10^{-4}$ of the impactor mass on Mars, and $2.3 \times 10^{-4}$ on Earth, although the fraction of it that would be in >3 m objects varies, from none for small impactors to nearly all for the largest. The ejected mass would then be distributed between destinations of the Solar System in accordance with our calculated probabilities, as in Tables 4 and 5. The impactor that created the Chicxulub crater and caused mass extinction on Earth ~ 65 Myr ago (Schulte *et al.* 2010) likely launched about $7 \times 10^{11}$ kg of rock into orbit, of which about 20,000 kg (~$7 \times 10^{-10}$ of the impactor's mass) could have reached Europa. For a single impact of this size, the probability of a fragment > 3 m reaching Europa is 0.6, and over 3.5 Ga we can expect one or two rocks to have made the journey due to impacts of this size.

Combining the above ejection rates with the rates of transfer, we estimate some tens of thousands of meteoroids from each planet should have been transferred to Jupiter, and several thousand from Earth should have hit Saturn. The transfer to these and the other planets are shown in Table 4.

Based on the simulations described above, we can approximate the likelihood of hitting Europa as a scaling factor times the probability of hitting Jupiter, where the scaling factor is derived from the shorter simulations described in the previous section. The expected transfer rates to the moons of the outer Solar System are shown in Table 5. We find expectation values of one to a few objects reaching Io, Europa, Ganymede, Callisto, Enceladus, and Titan from Earth since the LHB, and similar numbers for Jupiter's moons from Mars. Transfer from Mars to the moons of Saturn is less likely, but the upper limits we calculated do not completely rule out the possibility.

The above values are based on post-LHB impact rates. The LHB was a period of intense meteor impacts on the inner Solar System from 4.1 to 3.8 Gyr ago. Chapman *et al.* (2007) find that more than half of large impact craters date from the LHB, and Ivanov & Hartmann (2007) imply that LHB impacts may be even more numerous. This indicates that we can expect at least as many impacts to have occurred during the LHB as in the period of time since. Whether life was already present during this period is unknown, but if so, that could substantially raise the number of potentially life-bearing rock transfers between planets.

*4.4 Effects of returned material*
We observe that a large portion of material ejected returns to the planet of origin on short timescales of ≤1 Myr. Wells *et al.* (2003) studied the effect of returned material on restoring life after sterilizing impacts during the LHB. If life had arisen by this time, it may have been present on some of the many fragments ejected by the bombardment. The ejected fragments could have served as a refuge for life while the surface cooled enough to permit survival, then fallen back to re-seed life (Maher & Stevenson, 1988; Sleep & Zahnle, 1998). The fall-back rates in the first 3,000 years after impact were characterized by Wells *et al.* (2003). Our results show these rates over a longer timescale, but with a coarser grid -- our data was written in 1000-year timesteps.

We find that the cumulative probability of re-impact continues to increase substantially through the first few $10^5$ years, as shown in Fig. 4. We also note that Earth's higher gravity allows it to recapture far more of its ejecta than Mars can. This is despite the fact that we scaled our ejection velocities to the planet's escape velocity, resulting in Mars' ejecta having lower residual velocities than that from Earth.

*4.5 Discussion of uncertainties*
We are able to include estimates of the Poisson errors in the counts of meteoroids in the simulations, but we cannot quantify the possibility of systematic errors. Due to the chaotic nature of complex dynamics simulations, effects may have unexpected consequences – for example, strong unsampled resonances could exist which would result in skewed collision rates. We are aiming for simply an order-of-magnitude estimate of the transfer rates.

Additionally, if future studies revise the numbers on which our assumptions are based, our probabilities for transfer can be applied to the updated values. For a different estimate of the number of viable rocks ejected, the final numbers scale linearly. If the timescales for organism viability are found to be higher or lower than estimated here, then the number of viable rocks delivered within this time frame will respectively be higher or lower as well. However, due to the varying rates of transfer over time, the amount of variation is less predictable.

*4.6 Area for future study*
One additional aspect of physics that was not included in this study is the Yarkovsky effect (Bottke *et al.,* 2006). Objects from approximately 10 cm to 10 km will feel an extra propulsive force due to asymmetrically radiating more energy from the warm "afternoon" side of the object. The magnitude of this force will depend on the speed and angle of the object's rotation, and a population of ejected rocks will likely have a wide distribution of rotations. For many ejected objects, it will likely have a negligible effect, but for some subset, it may be significant and lead to faster inward decay or outward transfer, depending on the direction of rotation. This could be a fruitful area for future study, and we are composing a module to include this effect in MERCURY.

**5. Conclusions**
We find that transfer of rock capable of carrying life has likely occurred from both Earth and Mars to all of the terrestrial planets in the solar system and Jupiter, and transfer from Earth to Saturn is also probable. Additionally, we find smaller but significant probabilities of transfer to the moons of Jupiter and Saturn from Earth, and from Mars to the moons of Jupiter. These estimates are dependent on the number of rocks assumed ejected from the planets of origin. Our results indicate that transfer of life to these moons cannot be ruled out, and searches for life on these objects should keep in mind the necessity of determining whether life arose independently or descended from common ancestors to Earth life. Any life found there cannot be assumed to be of independent origin.

The probability of life surviving such a journey or finding a tenable environment on arrival is beyond the scope of our research. However, we note that studies of Titan, Europa, and Callisto all indicate significant liquid water oceans beneath the surface (Khurana et al., 1998; Lorenz et al., 2008). Europa currently presents the thinnest surface ice layer, providing less of a barrier for life to eventually find its way through, especially when considering the "chaos regions" that indicate recent partial melting. It appears regions of the ice sheet sometimes break into large chunks separated by liquid water, which later refreezes. Any meteorites lying on top of the ice sheet in a region when this occurs would stand a chance of falling through. Additionally, the moons are thought to have been significantly warmer in the not-too-distant past. Titan currently has a roughly 50 km thick crust, but the moon only cooled enough to form this shell after four billion years (Trobie *et al.,* 2006), before which it had only a few kilometers of methane clathrate over the surface, allowing a significant time in which life could have more easily penetrated into the liquid water ocean. Jupiter's moons are also believed to have been significantly warmer in the past, both due to residual heat of formation and their slow outward migration, making them previously subjected to stronger tidal heating from Jupiter.

   Ultimately, we conclude that the possibility of transfer of life from the inner Solar System to outer moons cannot be ruled out based on current knowledge. Any planned missions to search for life on Titan or the moons of Jupiter will have to consider whether any biological material found represents an independent origin, rather than another branch in the family tree populated by Earth life.

**Acknowledgements**
The authors thank the NASA Astrobiology Institute (NNA09DA76A) and the Penn State Astrobiology Research Center for their support.

**Author Disclosure Statement**
No competing financial interests exist.

**Table 1.** Total numbers and rates of transfer observed in our simulations by origin and destination. *No Mars-to-Saturn transfers were seen in the simulations, so we show the upper limit on the probability. The value is the probability for a single simulation transfer.

|         | Earth  | Earth (%)         | Mars   | Mars (%)          |
|---------|--------|-------------------|--------|-------------------|
| Orbit   | 17,484 | 40 ± 0.3          | 30,384 | 75 ± 0.4          |
| Sun     | 659    | 1.5 ± 0.06        | 508    | 1.3 ± 0.06        |
| Mercury | 159    | 0.37 ± 0.03       | 12     | 0.03 ± 0.009      |
| Venus   | 5,713  | 13 ± 0.2          | 617    | 1.5 ± 0.06        |
| Earth   | 17,201 | 40 ± 0.3          | 1,048  | 2.6 ± 0.08        |
| Mars    | 79     | 0.18 ± 0.02       | 6,362  | 16 ± 0.2          |
| Jupiter | 18     | 0.41 ± 0.01       | 16     | 0.04 ± 0.01       |
| Saturn  | 3      | 0.0069 ± 0.004    | 0      | < 0.0025*         |
| Ejected | 2,184  | 5 ± 0.1           | 1,553  | 3.8 ± 0.1         |
| Total   | 43,500 | 100               | 40,500 | 100               |

**Table 2.** Characterization of the transfer time distributions for each origin-destination pair. Time of first impact shows the minimum amount of time that was needed for an object to reach this destination after being ejected. Median impact times are given for both the 10 and 30 Myr simulations. The former is representative of the typical transfer time for possible viable organisms, while the latter more closely describes the overall distribution of ejecta.

|  | Destination | Time of first impact (yr) | 10 Myr Median (Myr) | 30 Myr Median (Myr) |
|---|---|---|---|---|
| from Earth | Sun | 1,014,612 | 7.1 | 12 |
|  | Mercury | 482,101 | 6.3 | 9.9 |
|  | Venus | 8,338 | 1.7 | 2.4 |
|  | Earth | 0.303 | 0.096 | 0.095 |
|  | Mars | 112,689 | 4.7 | 6.4 |
|  | Jupiter | 3,852,399 | 7.6 | 9.4 |
|  | Saturn | 8,008,387 | 8.8 | 8.8 |
|  | Ejected | 670,965 | 6.6 | 12 |
| from Mars | Sun | 1,737,473 | 7.5 | 15 |
|  | Mercury | 4,949,372 | 7.7 | 18 |
|  | Venus | 456,864 | 5.8 | 13 |
|  | Earth | 281,754 | 5.2 | 9.3 |
|  | Mars | 0.575 | 0.21 | 0.23 |
|  | Jupiter | 4,831,160 | 7.9 | 19 |
|  | Saturn | - | - | - |
|  | Ejected | 1,570,842 | 7.5 | 15 |

**Table 3.** Number of impacts per object in moon simulations.

| Body | Number | Body | Number |
|---:|---:|---:|---:|
| Jupiter | 533,776 | Saturn | 27,597 |
| Io | 73 | Enceladus | 2 |
| Europa | 36 | Titan | 3 |
| Ganymede | 40 | | |
| Callisto | 7 | | |

**Table 4.** Probabilities and estimated viable transfers to the planets in the Solar System. The 3.5 Gyr total is the number of transferred objects we expect. Mass transferred assumes each object is ~340,000 kg (i.e. a 3 m sphere with density of 3 g/cm³). No Mars-to-Saturn transfers were seen in the simulations, so we show the upper limits.

|  |  | Probability (%) | 3.5 Gyr Total | Mass (kg) transferred |
|---|---|---|---|---|
| from Earth | Orbit | 40 ± 0.3 | 80,000,000 ± 600,000 | $2.7 \times 10^{13}$ |
|  | Sun | 1.5 ± 0.06 | 3,000,000 ± 100,000 | $1.0 \times 10^{12}$ |
|  | Mercury | 0.37 ± 0.03 | 730,000 ± 60,000 | $2.5 \times 10^{11}$ |
|  | Venus | 13 ± 0.2 | 26,000,000 ± 300,000 | $8.9 \times 10^{12}$ |
|  | Earth | 40 ± 0.3 | 79,000,000 ± 600,000 | $2.7 \times 10^{13}$ |
|  | Mars | 0.18 ± 0.02 | 360,000 ± 40,000 | $1.2 \times 10^{11}$ |
|  | Jupiter | 0.041 ± 0.01 | 83,000 ± 20,000 | $2.8 \times 10^{10}$ |
|  | Saturn | 0.0069 ± 0.004 | 14,000 ± 8,000 | $4.7 \times 10^{9}$ |
|  | Ejected | 5 ± 0.1 | 10,000,000 ± 200,000 | $3.4 \times 10^{12}$ |
| from Mars | Orbit | 75 ± 0.4 | 600,000,000 ± 3,000,000 | $2.0 \times 10^{14}$ |
|  | Sun | 1.3 ± 0.06 | 10,000,000 ± 300,000 | $3.4 \times 10^{12}$ |
|  | Mercury | 0.03 ± 0.009 | 240,000 ± 50,000 | $8.1 \times 10^{10}$ |
|  | Venus | 1.5 ± 0.06 | 12,000,000 ± 400,000 | $4.1 \times 10^{12}$ |
|  | Earth | 2.6 ± 0.08 | 21,000,000 ± 500,000 | $7.0 \times 10^{12}$ |
|  | Mars | 16 ± 0.2 | 130,000,000 ± 1,000,000 | $4.3 \times 10^{13}$ |
|  | Jupiter | 0.04 ± 0.01 | 320,000 ± 60,000 | $1.1 \times 10^{11}$ |
|  | Saturn | < 0.0025 | < 20,000 | $< 6.8 \times 10^{9}$ |
|  | Ejected | 3.8 ± 0.1 | 31,000,000 ± 600,000 | $1.0 \times 10^{13}$ |

**Table 5.** Probabilities and estimated viable transfers over the past 3.5 Gyr to some of the moons in the outer Solar System. The impact ratio is the number of meteoroids that hit the planet (Jupiter or Saturn) for each one that hits the moon. For transfer from Mars to Saturn's moons, upper limits are used. The mass transfer rates are estimated by assuming that each transferred fragment is a sphere 3 m in diameter, with a density of 3 g/cm$^3$, i.e. a mass of 340,000 kg each.

|  | Destination | Impact Ratio | Probability (%) | 3.5 Gyr Total | Mass Transfer (kg) |
|---|---|---|---|---|---|
| from Earth | Io | 7,300 ± 900 | 5.7E-6 ± 7.0E-7 | 10 ± 1 | 3,800,000 |
|  | Europa | 15,000 ± 2,000 | 2.8E-6 ± 5.0E-7 | 6 ± 0.9 | 1,900,000 |
|  | Ganymede | 13,000 ± 2,000 | 3.1E-6 ± 5.0E-7 | 6 ± 1 | 2,100,000 |
|  | Callisto | 76,000 ± 30,000 | 5.4E-7 ± 2.0E-7 | 1 ± 0.4 | 370,000 |
|  | Enceladus | 14,000 ± 10,000 | 5.0E-7 ± 4.0E-7 | 3 ± 2 | 1,000,000 |
|  | Titan | 9,200 ± 5,000 | 7.5E-7 ± 4.0E-7 | 4 ± 3 | 1,500,000 |
| from Mars | Io | 7,300 ± 900 | 5.4E-6 ± 6.0E-7 | 10 ± 1 | 3,700,000 |
|  | Europa | 15,000 ± 2,000 | 2.7E-6 ± 4.0E-7 | 5 ± 0.9 | 1,800,000 |
|  | Ganymede | 13,000 ± 2,000 | 3.0E-6 ± 5.0E-7 | 6 ± 0.9 | 2,000,000 |
|  | Callisto | 76,000 ± 30,000 | 5.2E-7 ± 2.0E-7 | 1 ± 0.4 | 350,000 |
|  | Enceladus | 14,000 ± 10,000 | < 1.8E-7 | < 1.4 | < 490,000 |
|  | Titan | 9,200 ± 5,000 | < 2.7E-7 | < 2.1 | < 730,000 |

**Figure 1.** Histograms showing the distribution in impact time for each location, in million-year bins.

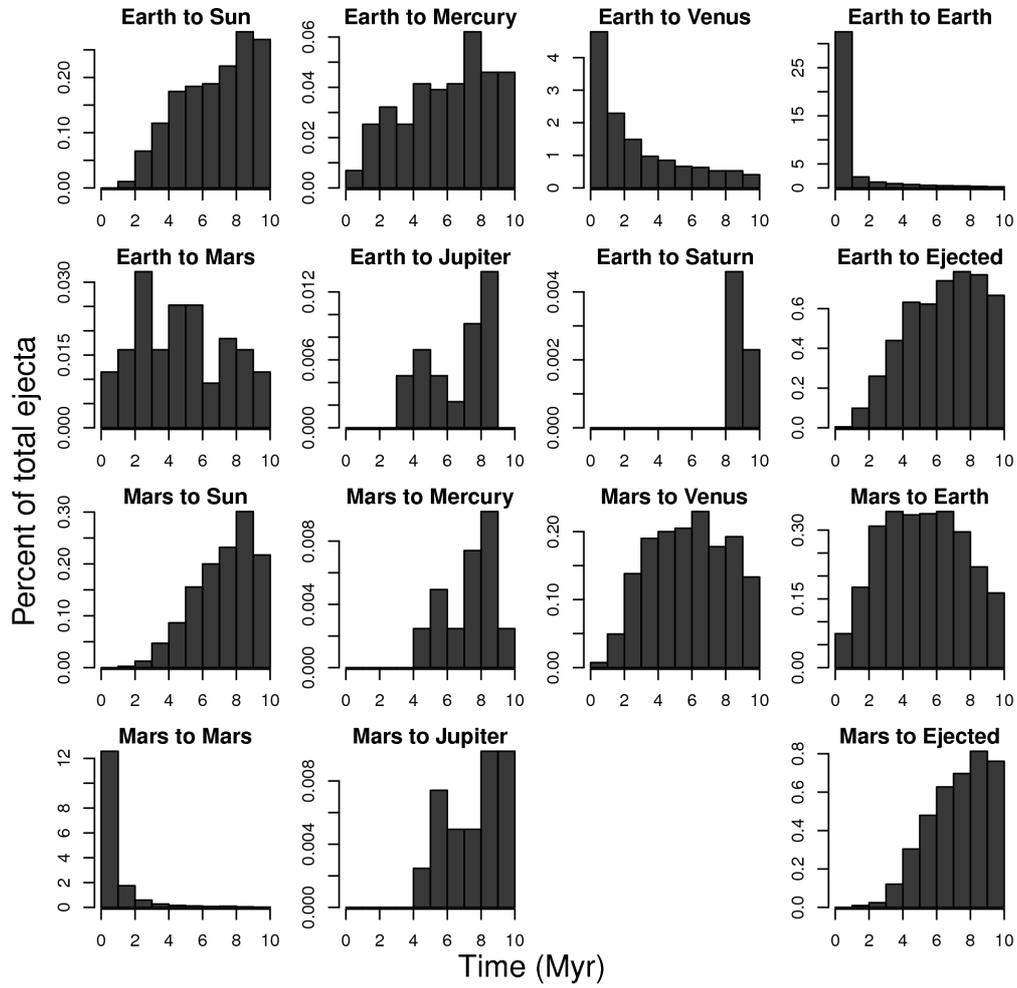

**Figure 2.** The results of our moon simulations, plotted against different predictors. The lettered plot symbols identify Io, Europa, Ganymede, Callisto, and Titan by their first letter, and Enceladus is represented with an "N." (**a**) The Liouville theorem prediction is represented by the black line. "A" is the cross-section area, "$F_g$" is focusing factor, and subscripts "m" and "p" indicate moon and planet, respectively. (**b**) A useful empirical correlation was observed between the moon-planet collision ratio and a combination of the moon and planet parameters. The x-axis is the gravitational focusing factor of the moon divided by that of the planet times the planet's Hill sphere radius, divided by the square of the moon's semimajor axis. The black line is a fit to the points, with a slope of $4.8 \times 10^9$ and intercept $1.3 \times 10^{-6}$.

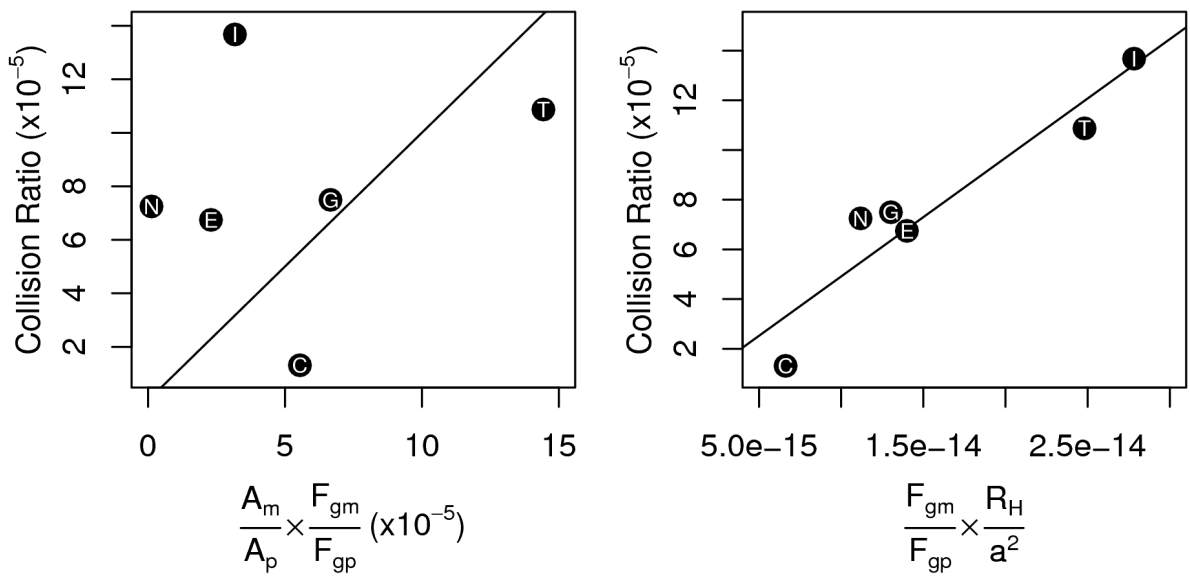

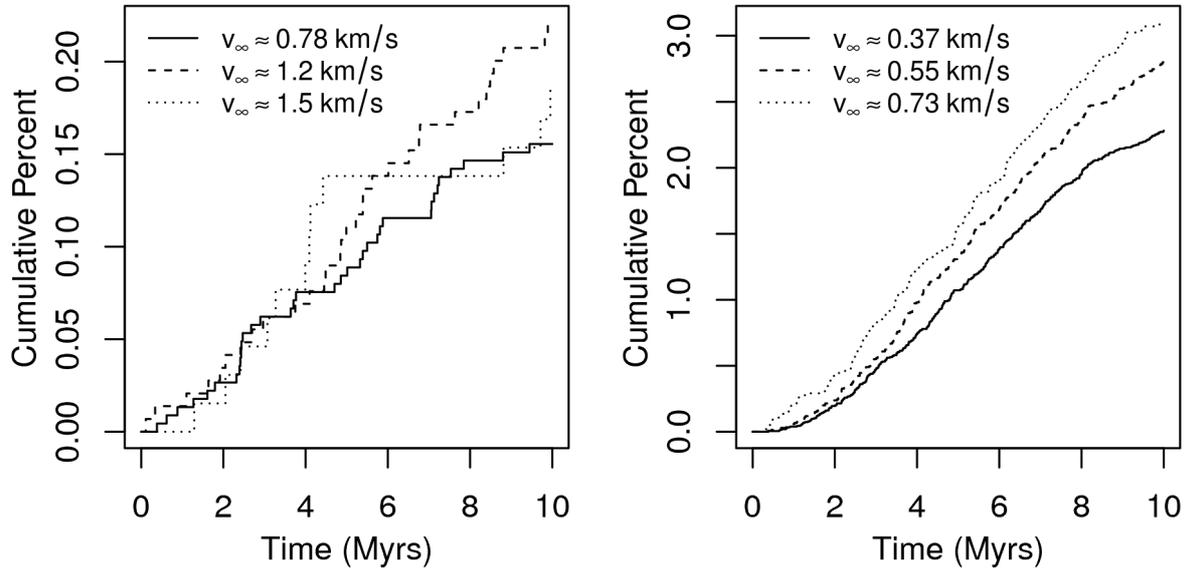

**Figure 3.** Cumulative collision rates for transfer (**a**) from Earth to Mars and (**b**) from Mars to Earth, separated by ejection velocity.

**Figure 4.** The percentage of ejected material returned to its planet of origin over the first million years after impact.

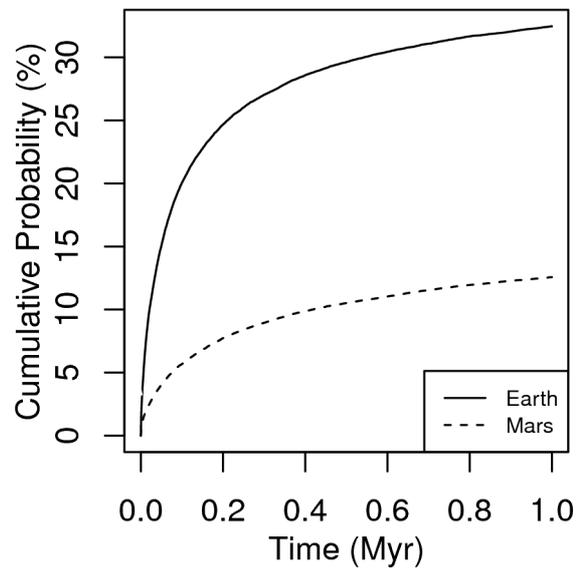

**Bibliography**

Armstrong, J. C., Wells, L. E., and Gonzalez, G. (2002) Rummaging through Earth's attic for remains of ancient life. *Icarus* 160:183-196.

Barber, D. J. and Scott, E. R. D. (2002) Origin of supposedly biogenic magnetite in the Martian Meteorite Allan Hills 84001. *PNAS* 99 (10):6556-6561.

Benzerara, K., Menguy, N., Guyot, F., Dominici, C., and Gillet, P. (2003) Nanobacteria-like calcite single crystals at the surface of the Tataouine meteorite. *PNAS* 100:7438–42.

Bogard, D. D. and Johnson, P. (1983). Martian gases in an Antarctic meteorite?. *Science*, *221*(4611), 651-654.

Bottke Jr., W. F., Vokrouhlický, D., Rubincam, D. P., and Nesvorný, D. (2006) The Yarkovsky and YORP effects: Implications for asteroid dynamics. *Annual Review of Earth and Planetary Science* 2006.34:157-191.

Carr, R. H., Grady, M. M., Wright, I. P., and Pillinger, C. T. (1985). Martian atmospheric carbon dioxide and weathering products in SNC meteorites. *Nature*, *314*(6008), 248-250.

Chambers, J. E. (1999) A hybrid symplectic integrator that permits close encounters between massive bodies. *MNRAS* 304 (4):793-799.

Chapman, C. R., Cohen, B. A., and Grinspoon, D. H. (2007) What are the real constraints on the existence and magnitude of the Late Heavy Bombardment? *Icarus* 189(1):233-245.

Cisar, J.O., Xu, D.Q., Thompson, J., Swaim, W., Hu, L., and Kopecko, D.J. (2000) An alternative interpretation of nanobacteria-induced biomineralization. *PNAS* 9:11511–15.

Dones, L., Gladman, B., Melosh, H. J., Tonks, W. B., Levison, H. F., and Duncan, M. (1999) Dynamical lifetimes and final fates of small bodies: orbit integrations vs. Öpik calculations. *Icarus* 142:509-524.

Gladman, B. (1997) Destination: Earth. Martian meteorite delivery. *Icarus* 130:228–246.

Gladman, B. and Burns, J.A. (1996) Mars meteorite transfer: Simulation.

*Science (Letters)* 274(5285):161–162.

Gladman, B., Burns, J.A., Duncan, M., Lee, P., and Levison, H.F. (1996) The exchange of impact ejecta between terrestrial planets. *Science* 271(5254):1387–1392.

Gladman, B., Dones, L., Levison, H.F., and Burns, J.A. (2005) Impact seeding and reseeding in the inner Solar System. *Astrobiology* 5(4):483–496.

Gladman, B., Dones, L., Levison, H.F., Burns, J.A., and Gallant, J. (2006) Meteoroid transfer to Europa and Titan. In *Lunar and Planetary Science XXXVII*.

Golden, D. C., Ming, D. W., Morris, R. V., Brearley, A. J., Lauer, H. V., Treiman, A. H., Zolensky, M. E. , Schwandt, C. S., Lofgren, G. E., and McKay, G. A. (2004) Evidence for exclusively inorganic formation of magnetite in Martian meteorite ALH84001. *American Mineralogist* 89(5-6): 681-695.

Halevy, I., Fischer, W. W., and Eiler, J. M. (2011) Carbonates in the Martian meteorite Allan Hills 84001 formed at 18±4° C in a near-surface aqueous environment. *PNAS 108*(41):16895-16899.

Hut, P., and Bahcall, J. N. (1983) Binary-single star scattering. I-Numerical experiments for equal masses. *The Astrophysical Journal, 268*, 319-341.

Irving, A. (2004) New Martian meteorite found in Algeria: NEA 3171. http://www2.jpl.nasa.gov/snc/nwa3171.html.

Jet Propulsion Laboratory (JPL). HORIZONS system. URL http://ssd.jpl.nasa.gov/?horizons.

Khurana, K. K., Kivelson, M. G., Stevenson, D. J., Schubert, G., Russell, C. T., Walker, R. J., and Polanskey, C. (1998) Induced magnetic fields as evidence for subsurface oceans in Europa and Callisto. *Nature, 395*(6704), 777-780.

Lorenz, R. D., Stiles, B. W., Kirk, R. L., Allison, M. D., del Marmo, P. P., Iess, L., Lunine, J. I., Ostro, S. J., and Hensley, S. (2008) Titan's rotation reveals an internal ocean and changing zonal winds. *Science, 319*(5870), 1649-1651.

Maher, K. A. and Stevenson, D. J. (1988) Impact frustration of the origin of life. *Nature* 331(6157):612-614.

Martel, Jan; Young, David; Peng, Hsin-Hsin; Wu, Cheng-Yeu; and Young, John D. (2012) *Biomimetic Properties of Minerals and the Search for Life in the Martian Meteorite ALH84001*. Annual Review of Earth and Planetary Sciences, 40(1):167-193

McKay, C.P., Mancinelli, R.L., Stoker, C.R., and Wharton Jr, R.A. (1992) The possibility of life on Mars during a water-rich past. Mars, pages 1234–1245. http://adsabs.harvard.edu/abs/1992mars.book.1234M.

McKay. D.S., Gibson Jr., E.K., Thomas-Keprta, K.L., Vali, H., Romanek, C.S., Clemett, S.J., Chillier, X.D.F, Maechling, C.R., and Zare, R.N. (1996) Search for past life on Mars: Possible relic biogenic activity in Martian meteorite ALH84001. Science, 273(5277):924–930.

Melosh, H. J. (1988). The rocky road to panspermia. *Nature*, *332*(6166), 687.

Melosh, H. J. (2003) Exchange of meteorites (and life?) between stellar systems. *Astrobiology,* 3(1):207-215.

Melosh, H.J. and Tonks, W.B.. (1993) Swapping rocks: Ejection and exchange of surface material among the terrestrial planets. Meteoritics, 28(3):398.

Mileikowsky, C., Cucinotta, F.A., Wilson, J.W., Gladman, B., Horneck, G., Lindegren,L. , Melosh, J., Rickman, H., Valtonen, M., and Zheng, J.Q. (2000) Natural transfer of microbes in space; 1. From Mars to Earth and Earth to Mars. Icarus, 145(2): 391–427.F

Nyquist, L. E., Bogard, D. D., Shih, C. Y., Greshake, A., Stöffler, D., and Eugster, O. (2001) Ages and geologic histories of Martian meteorites. *Space Science Reviews* 96(1-4):105-164.

R Core Team. (2013) R: A Language and Environment for Statistical Computing. R Foundation for Statistical Computing, Vienna, Austria. http://www.R-project.org

Reyes-Ruiz, M., Chavez, C.E., Hernandez, M.S., Vazquez, R., Aceves, H., and Nunez, P.G. (2012) Dynamics of escaping Earth ejecta and their collision probability  with different Solar System bodies. *Icarus* 220(2):777–786.

Schulte, P., Alegret, L., Arenillas, I., Arz, J. A., Barton, P. J., Bown, P. R., Bralower, T.J., Christeson, G.L., Claeys, P., Cockell, C.S., Collins, G.S., Deutsch, A., Goldin, T.J., Goto, K., Grajales-Nishimura, J.M., Grieve, R.A.F.,

Gulick, S.P.S., Johnson, K.R., Kiessling, W., Koeberl, C., Kring, D.A., MacLeod, K.G., Matsui, T., Melosh, J., Montanari, A., Morgan, J.V., Neal, C.R., Nichols, C.J., Norris, R.D., Pierazzo, E., Ravizza, G., Rebolledo-Vieyra, M., Reimold, W.U., Robin, E., Salge, T., Speijer, R.P., Sweet, A.R., Urrutia-Fucugauchi, J., Vajda, V., Whalen, M. T., and Willumsen, P. S. (2010) The Chicxulub asteroid impact and mass extinction at the Cretaceous-Paleogene boundary. *Science*, *327*(5970), 1214-1218.

Sleep, N. H. and Zahnle, K. (1998) Refugia from asteroid impacts on early Mars and the early Earth. *Journal of Geophysical Research: Planets (1991–2012)*, *103*(E12), 28529-28544.

The Meteoritical Society. (2012) Meteoritical bulletin database, http://www.lpi.usra.edu/meteor/metbull.php.

Tobias, C. A., & Todd, P. (1974). Radiation and molecular and biological evolution. *Space radiation biology and related topics. New York, Academic Press, Inc.*, 197-255.

Thomas-Keprta, K.L., Bazylinski, D.A., Kirschvink, J.L., Clemett, S.J., McKay, D.S., Wentworth, S.J., Vali, H., Gibson Jr, E.K., and Romanek, C.S. (2000) Elongated prismatic magnetite crystals in ALH84001 carbonate globules: potential Martian magnetofossils. *Geochim Cosmochim Acta* 64(23):4049–4081.

Trobie, G., Lunine, J.I., and Sotin, C. (2006) Episodic outgassing as the origin of atmospheric methane on Titan. *Nature* 440(7080):61–64.

Wells, L. E., Armstrong, J. C., Gonzalez, G. (2003) Reseeding of early Earth by impacts of returning ejecta during the late heavy bombardment. *Icarus* 162:38-46.

**APPENDIX: Linear modeling**

Our data include both continuous (numerical) and categorical (non-numerical) variables, which require different treatments when fitting. When both predictor and response are continuous, we fit a line of the form $Y=\beta_1*X+\beta_0$, where $\beta_0$ is the intercept of the line, and $X$ and $\beta_1$ are the predictor variable and its corresponding slope. In the case of multiple predictors, $X$ is replaced with the predictors $X_1, X_2$, etc., each of which will have its own slope $\beta_1, \beta_2$, etc. If the predictor is categorical, $\beta$ simply becomes the difference between the average response values for the two categories.

For a categorical response variable such as destination (where the value may be collision with the Sun or other planets, ejection from the system, or remaining in orbit), the meaning of the actual fit parameters is somewhat less straightforward. In this case, $\beta$ represents the change in the natural log of the odds when the predictor changes from one category to the other. The model treats the first category as the baseline, and uses a binomial model for each of the other categories paired with the baseline. For example, remaining in orbit (i.e. destination = orbit) is the baseline, and each of the other destinations is evaluated against the baseline. For the case of comparing "destination = Earth" versus "destination = orbit", the baseline value (orbit) is designated a negative result, with a value of 0, and Earth is positive and has a value of 1. The probability of impacting Earth compared to that of staying in orbit, when the planet of origin is Earth, is then

$$P_{01} = N_{E2E}/(N_{E2E}+N_{E2O}).$$

where $N_{E2E}$ and $N_{E2O}$ are the number of objects from Earth that hit Earth or stay in orbit, respectively. Similarly, for the probability of staying in orbit,

$$P_{00} = N_{E2O}/(N_{E2E}+N_{E2O}).$$

Correspondingly, the probability for objects from Mars of impacting Earth compared to staying in orbit is

$$P_{11} = N_{M2O}/(N_{M2E}+N_{M2O})$$

and so on. The odds are defined as the ratio of probabilities of a positive result over a negative one, so the odds of hitting Earth for an object from Earth, when comparing the destinations of Earth versus orbit, would be $P_{01}/P_{00}$ which reduces to $N_{E2E}/N_{E2O}$, and the odds for an object from Mars would be $P_{11}/P_{10}=N_{M2E}/N_{M2O}$. The parameter being fit by the model, $\beta_1$, is the change in the natural log of the odds when the predictor's value increases by one or, for a categorical variable like planet of origin, changes from the first to the second category, i.e.

$$\beta_1 = ln(N_{M2E}/N_{M2O}) - ln(N_{E2E}/N_{E2O})$$
$$\beta_1 = ln((N_{M2E}/N_{M2O})/(N_{E2E}/N_{E2O}))$$

Equivalently, the odds of hitting Earth for an object from Mars would be $e^{\beta_1}$ times the odds for an object from Earth.

$$(N_{M2E}/N_{M2O}) = (N_{E2E}/N_{E2O}) e^{\beta_1}$$

Thus, a positive $\beta$ implies the odds increase as you go from the first to the second category, and negative implies a decrease.